# TITLE:

# Stress-enhanced ion diffusion at the vicinity of a crack tip as evidenced by atomic force microscopy in silicate glasses.

AUTHORS:

F. Célarié[1], M. Ciccotti[2], C. Marlière[3*]

(1) Institute of Non-Metallic Materials, Clausthal University of Technology, Clausthal-Zellerfeld, Germany

(2) Laboratoire des Colloïdes, Verres et Nanomatériaux (L.C.V.N.), University Montpellier II, CNRS, France.

(3) Géosciences Montpellier (G.M.), University Montpellier II, CNRS, France.

---

[*] Corresponding author. Email address : christian.marliere@univ-montp2.fr

# **ABSTRACT**

The slow advance of a crack in sodo-silicate glasses was studied at nanometer scale by in-situ and real-time atomic force microscopy (AFM) in a well-controlled atmosphere. An enhanced diffusion of sodium ions in the stress-gradient field at the sub-micrometric vicinity of the crack tip was revealed through several effects: growth of nodules in AFM height images, changes in the AFM tip–sample energy dissipation. The nodules patterns revealed a dewetting phenomenon evidenced by "breath figures". Complementary chemical micro-analyses were done. These experimental results were explained by a two-step process: i) a fast migration (typical time: few milliseconds) of sodium ions towards the fracture surfaces as proposed by Langford et al. [J. Mat. Res. 6 (1991) 1358], ii) a slow backwards diffusion of the cations as evidenced in these AFM experiments (typical time: few minutes). Measurements of the diffusion coefficient of that relaxing process were done at room temperature. Our results strengthen the theoretical concept of a near-surface structural relaxation due to the stress-gradient at the vicinity of the crack tip. The inhomogeneous migration of sodium ions might be a direct experimental evidence of the presence of sodium-rich channels in the silicate structure [A. Meyer et al., Phys. Rev. Let. 93 (2004) 027801].

## PACS INDEXING CODES

| | |
|---|---|
| 62.20.Mk | Fatigue, brittleness, fracture, and cracks |
| 66.30.-h | Diffusion in solids |
| 68.03.Cd | Surface tension and related phenomena |
| 68.37.Ps | Atomic force microscopy (AFM) |

# I. INTRODUCTION

The study of mechanisms of rupture in so-called brittle materials is still of great importance both for knowledge of basic mechanisms of bond(s) breaking and for practical reasons such as understanding and possibly correcting the degradation of nuclear waste containers or optical glass fibers or predicting rupture in the upper Earth's crust as well. However and despite of recent tremendous research work discussions about the real processes and their characteristic length scales are still intense.

One important question is related to know whether the fundamental mechanism of brittle fracture at the crack tip in silicate glasses (at temperature far below their glass transition temperature) is related to the one of following processes: i) sequential atomic bond rupture (the so-called *brittle* process) with a characteristics length scale related to rupture process in the order of few tenths of nanometers or ii) a so-called *plastic-like* deformation in which the zone affected by bond rupture or atoms/ions transfer is in range of up to few tens of nanometers?

From theoretical arguments by Kelly et al. [1], Hillig [2], and Rice and Thomson [3], Lawn et al. [4] argued that covalent or ionic solids as silicate glasses will generally support perfectly brittle cracks. They break by sequential bond rupture on the opposite to metallic solids that can deform plastically at crack tips for both fracture regimes: high speed fracture and very slow cracking, the so-called stress corrosion regime (SCR). Recently Guin et al. [5, 6] studied at nanometer scale by means of atomic force microscope (AFM) the traces left by the advance of the crack line through silica or soda-lime glasses on both fracture surfaces. This was done in the SCR in liquid water and for fast crack in aqueous gaseous atmosphere. That *post-mortem* study allows these authors to conclude that crack tip appeared atomically sharp and a completely brittle fracture process occurs. In case of soda-lime glasses in which mobile sodium ions are present Fett et al. [7] concluded from their observations that ion exchange between hydronium ($H_3O^+$) ions in the solution and sodium ($Na^+$) ions in the glass gives rise to compressive stresses and is responsible for the occurrence of a fatigue limit in glass, as already suggested by Bunker et al. [8-9]. For these authors the crack tip should remain atomically sharp at the fatigue limit. That idea well fits the classical theory due to Barenblatt [10] who predicted that the rounding of the crack end should be suppressed under the action of short-range forces and consequently replaced by a sharp cusp.

However that concept is a subject of strong scientific debate. Plasticity (resulting in any permanent deformation on distances higher than a few interatomic bonds at least) has been observed indeed for cracks in glass under compression (e.g., indentation) [11]. Transmission electron microscopy (TEM) study [12] of crack into thin layers of silica glass indicated a crack tip radius as large as 1.5 nm. Furthermore crack tip rounding was observed after soaking the glass in water at 90°C and was attributed to reprecipitation of silica at the crack tip. That effect was considered as a plausible explanation to the increase of mechanical strength of abraded soda-lime glasses as observed after soaking in water or by annealing. On the opposite, the experimental observation of a rounded crack tip was explained by Lawn et al. [4] as due to the removal of tensile stress that existed near the crack tip considered as being atomically sharp. More recently Guilloteau et al. [13] studied the vicinity of a crack tip propagating in tensile mode through soda-lime glass. They noticed a clear breakdown of the 1/2 power law for distances to the crack tip lower than 50 nanometers. They noted too that the width of the crack (as measured behind its tip) is much larger –and in a permanent way -than the prediction of the linear elasticity. From these observations Guilloteau et al. deduced [13] that a non-linear behaviour takes place in the close vicinity of the crack tip. These results were recently confirmed in case of alumino-silicate [14] or silica [15] glasses. Furthermore by real-time and in-situ AFM observation of slow propagation of a tensile crack in gaseous aqueous atmosphere (mixture of nitrogen and water vapor only) Célarié et al. [14] reported the formation of damage cavities, 20 nm wide by 5 nm deep ahead of the crack tip, and attributed crack propagation in glass to the formation and coalescence of these cavities. It was proposed that their presence explains the departure from linear elasticity observed in the vicinity of a crack tip in glass. These observations are in agreement with a scenario predicted by molecular dynamics simulations in case of fast cracks propagating in vacuum [16, 17]. Recently, Xi *et al.* [18] showed that so-called brittle metallic glasses also fracture by cavity growth and coalescence. It must be noted too that a ductile-like opening mode of fracture by pore growth and coalescence has been independently invoked for so-called 'brittle' rock failure [19].

Tomozawa et al. [20] proposed that surface structural relaxation of pure silica glass may be a possible mechanism of the mechanical fatigue. It is well known that glasses are in a metastable state and thus underwent bulk structural relaxation. The kinetics of such relaxation are extremely slow at a temperature far below glass transition temperature [21]. The results as obtained by IR reflection method showed that the kinetics of surface structural relaxation is much faster than in the bulk [22] and is greatly enhanced by the presence of water vapor [23] and/or applied tensile stress [22]. An experimental evidence of water entry into silica glass during fracture was brought by Tomozawa et al. [24]. Tomozawa [25] has shown that the slow crack growth velocity is proportional to the diffusion coefficient of water, greatly accelerated by the tensile stress near the crack tip. The entry of water inside of a silica sample under tensile stress is in the range of few tens of nanometers at room temperature in time period of typically few minutes [26]. Such relaxation effects with probable mass transfer via atoms or ions diffusion could be at the origin of the observed damage cavities [14].

In order to try and clarify the debate we aim to study the sub-micrometric vicinity of the tip of a crack running in silicate glass with mobile ions (as sodium cations) and check for possible chemical mass transfer at nanometer range. Soda-lime glasses, despite of their chemical complexity, were good candidates as i) samples with a high homogeneity of composition (a mandatory condition for the double cleavage drilled compression –DCDC- samples we used) can be easily obtained and ii) such local ionic migration was already suspected in complex conditions [27-28].

Diffusion of alkali elements near the surface of silicate glasses is a phenomenon already well documented [29]. For instance electron [30] or ionic [31] irradiation of sodo silicate glasses causes a migration of sodium ions inside the bulk material. On the opposite a X-ray irradiation induces the increase of the Na surface concentration [32]. Migration of the mobile Na ions under a bending stress from zones in compression to regions of tension was observed by measurement of electrical current flowing through the soda-lime silicate sample [33]. During rapid (dynamic) fracture of sodium silicate glasses in vacuum emission of atomic sodium was observed by quadrupole mass spectroscopy and surface ionization techniques. The sodium emission occurs few ms (3-6ms) after fracture event and decays over tens of milliseconds [34].

This paper will now detail results obtained by AFM studies and ex-situ complementary experiments. The AFM experiments were done on samples under stress, during the very slow propagation of the crack. Thanks to wetting/dewetting phenomena in the vicinity of the crack tip, we clearly evidenced a slow diffusion process related to the stress gradient. The source of the diffusion species are shown to be located at the nanometer vicinity of the crack surfaces. Measurement of diffusion coefficient was done at room temperature. These results are interpreted as a two-step process: the stress field at the crack tip causes a fast migration (characteristic time in the range of milliseconds) of sodium element towards the crack surfaces in a similar manner as in ref [34]. In a second step the relaxation of that out-of equilibrium enrichment in sodium is relaxed thanks to a slow diffusion process (characteristic time equal to few minutes) as evidenced by the AFM data.

## II. EXPERIMENTAL SETUP

We used for this work soda-lime glasses[1] as prepared from commercial plate. The composition of the glass by molar fraction (%) is as follows: $SiO_2$ : 70.9; $Na_2O$ : 13.2 ; CaO : 10; MgO : 5.4 ;

---

[1] From Saint-Gobain company.

$Al_2O_3$ : 0.4;, $K_2O$ : 0.1. The samples were cut from the same piece of that soda-lime silicate glass. A thermal treatment (530°C) was done before fracture experiment in order to remove residual stresses. Fracture is performed on DCDC (double cleavage drilled compression) [35], parallepipedic (4mmx4mmx40mm) samples designed with a cylindrical hole (radius a=0.5 mm) drilled at the center of two parallel 4mmx40mm surfaces and perpendicularly to them. The hole axis defines the Z-direction. The X axis – what will be called the "crack direction"- (resp. Y axis – what will be called the "opening direction") is parallel to the 40mm (resp. 4 mm) side of the 4mmx40mm surfaces. Several cases of surface preparation were used for AFM experiments: we worked with both the "genuine" surfaces (in contact with air or with tin bath during cooling of glass melt) without any further polishing or with bulk samples in which the 4×40 $mm^2$ surfaces were optically polished. In all cases the measured RMS roughness was lower than 0.25 nm for a 10×10 $\mu m^2$ scan-size. No change in presented results was detected versus the nature of the studied surface. Samples were carefully cleaned by a slightly basic detergent in ultra-sound bath, then rinsed with ultra-pure water before being dried by pure nitrogen flow. They were then placed between the jaws of the stress machine and all the experimental set-up was inside a chamber with controlled atmosphere. A compressive load was applied perpendicularly to the 4mmx4mm surfaces [36] by a slow constant displacement (0.02 $mm.min^{-1}$) of the jaws of the compression machine. When the crack was launched the jaws were then blocked. The crack front propagates in pure opening mode (mode I) along the X-axis in the symmetry plane of the sample parallel to the (X, Z) plane. In this geometry, the stress intensity factor (SIF) is

$$K_I = \sigma a^{1/2}/(0.375c/a+2) \quad (1)$$

where *c* is the length of the crack and *a* is the radius of the hole in the DCDC specimen. The crack motion within the (X, Y) surface is monitored by our experimental system combining optical microscopy and Atomic Force Microscopy AFM (Dimension 3100, Nanoscope IIIa, Digital Instruments). For crack velocity ranging from $10^{-6}$ to $10^{-9}$ $m.s^{-1}$ we use the optical image (with a spatial resolution of 5 μm) and for lowest velocity we performed measurements with AFM in an amplitude resonant mode ("tapping" mode) with commercial Si/SiO2 cantilever (spring constant ~ 35 N/m) and integrated tip (tip radius around 10nm). These fracture experiments are performed at a constant temperature of 21.0 ± 1°C in a leak-proof chamber under an atmosphere composed of pure nitrogen and water vapour (preceding by a careful out-gassing). The relative humidity (RH) can be controlled from 5% to 75% with an accuracy of about 1%. It must be noted that we measured the mean (macroscopic) humidity rate inside the glove-box. That value may be different from that in the surroundings of the crack. Three different values of humidity rate were mainly studied: 45%RH, 48%RH and 58%RH.

After the AFM experiments, samples removed from the mechanical stage were immediately transferred to a primary vacuum chamber maintained at a pressure of about $10^3$ Pa before further external investigations. Brief exposure (<1min) to ambient atmosphere was unavoidable during this transfer. The samples were not completely broken and (X,Y) surfaces were studied. Surface was overlaid with platinum layer (2.5 nm in thickness) to avoid electrical charging during analyses. Two main chemical method of surface analyses were used:

- Electron Probe Micro-Analysis (EPMA) on a Cameca-SX100 equipped with Wavelength Dispersive X-ray Spectroscopy (WDS) system, done at "Service Commun Microsonde Sud" at University Montpellier II.

- Secondary Ions Mass Spectroscopy analyses (SIMS) were performed at "Service Commun Microsonde Sud" at University Montpellier II on a IMS4F (Cameca). The primary ions are negative oxygen ions, accelerated with 12.5 kV. A current of 6 nA with a beam focused on a zone of approximately 25 μm in diameter and sweeping a total area of 150μmx150μm was used for firstly removing the platinum layer and for further sputtering. For element analysis a primary current of approximately 1 pA was focused in a 1 μm$^2$ beam and swept a zone of approximately 50μmx50μm. The secondary ions were extracted with an acceleration of 4,5 kV and analyzed in energy and mass by an electron multiplier working in counting mode.

## III. RESULTS

### III.A. AFM experiments: evidence of a diffusion process.

*III.A.1.  Influence of crack speed on quadratic profile of the diffusion front.*

It must be noted that the phenomena described below were not detectable when relative humidity rate was lower than 30%. A typical topographical AFM frame in the neighborhood of the crack tip, for $K_I$ = 0.41 MPa.m$^{1/2}$, v = 1.5 nm.s$^{-1}$ and RH = 45 %, is presented in figure 1. It clearly reveals many nodules typically 200 nm in diameter for the largest (respectively 10nm for the smallest) and 20 nm in height for the highest (resp. 2 nm). It must be noted that the lower limit in diameter ( 10 nm) for the smallest, is due to the limited lateral resolution caused by the rounded apex of AFM tip (curvature radius around 10nm). These nodules nucleate and grow just around the crack lips. As we see in figure 1.b, ahead of the crack (on the right side of the AFM frame) they are not visible: the RMS roughness on the image section where nodules are visible is equal to 4nm ( as calculated on a 8μm$^2$ surface): this value is much higher than the corresponding one related to the "flat" zone in front of the crack tip. In that case the RMS roughness is equal to 0.38 nm as calculated on a rectangular area of 10μm$^2$. This value is similar to that obtained in figure 1.a, taken 30 minutes before the crack reached the region under AFM investigation: 0.40 nm as calculated on an 25μm$^2$ area. These values are typical of surfaces of silicate glasses.

It must be emphasized that, in the same conditions of humidity and mean crack speed, this phenomenon was not observed with silica samples [15] as well as with lithium alumino-silicate glasses[1] [14]: RMS roughness, as calculated for instance on an 1μm$^2$ area of lithium alumino-silicate glass, is equal to 0.20nm for the case (Fig 2.a) where crack did not reach the region of AFM investigation and 0.36nm in the opposite situation (Fig.2.b): these values are near of that measured on soda-lime glass in the "non-contaminated" (no nodules) zone as expected.

A preliminary remark has to be done. For a given sequence of AFM frames, such as that in figure 1, the crack speed can be considered as constant. By the experimental knowledge of relation between $K_I$ and $\log(v_c)$ [37] we indeed estimated that a crack advance of 7μm (for a crack length of 4,5mm) corresponds to a relative variation of crack speed of 1%. That value is below the relative uncertainty due to residual AFM drifts (estimated to 3%). So the XX' axis (along direction propagation of crack; see figure 1) can be rescaled to a time axis.

The contour delimiting the no-nodules zone and the contaminated region (what will be called "diffusion front") has a parabolic profile as it can be seen in figure 3 (white continuous line). This contour, delimiting the zone of the nucleation of nodules corresponds to following equation:

$$(x - x_0) = B \cdot (y - y_0)^2 \qquad (3)$$

Due to the almost constant crack speed during acquisition of an AFM image or a sequence of AFM images, the X-axis can be seen as a time axis and following relation can be written:

$$(x - x_0) = v_c \cdot (t - t_0) \qquad (4)$$

where $v_c$ is the mean crack speed. Please note that the typical AFM scan rate along X axis (slow scan axis) was chosen to be equal to at least twenty times greater than $v_c$.

Thus there is a quadratic relation between $y$ and $t$ as in a standard diffusion process in which the source of diffusing species is located on crack surfaces ($y = 0$ represents the crack lips in the 2D representation provided by AFM techniques). Classical theories of diffusion [38] predict indeed that the local concentration of the diffusion species is a function of variable $u$ with

$$u = (y - y_0)/(2 \cdot \sqrt{D_{eff} \cdot (t - t_0)}) \qquad (5)$$

where $D_{eff}$ is an effective diffusion coefficient.

---

[1] $SiO_2$ mol 70% ; $Al_2O_3$ mol 20%; $Li_2O$ mol 6%; $TiO_2$ mol 4%

The presence of the diffusing species is evidenced by these AFM experiments in height mode providing the local height of the compound built from the diffusing species is higher than the threshold of detection ( estimated to 0.07nm). That threshold is related to a given value, $u_o$, of $u$. From (3), (4) and (5) and by writing that $u = u_o$ at the diffusion front line we can estimate the value of $D_{eff}$ from relation:

$$D_{eff} = v_c / (4 \cdot B \cdot u_0^2) \qquad (6)$$

To estimate the value of $D_{eff}$ we proceed as following: the contour of diffusion front is determined from AFM frames; then a second order polynomial is fitted to experimental data by a least-squares method (figure 4). A typical fitting result is the solid line in figure 4 ($v_c$ = 1.5 nm/s). Good fits were obtained for all experiments based on height AFM frames ( $v_c$ values between 0.1nm/s to 3nm/s). However when crack speed increases nodules cannot be detected by AFM in *height* mode as it is seen in figure 5.a where $v_c$ = 7 nm/s. The increase in height of nodules between two successive AFM images occurring during the propagation of crack through the glass is indeed lower than the vertical AFM sensitivity. However the phase signal, sensitive to variation of the changes in the tip–sample energy dissipation (mainly due to modifications of adhesion properties; see detailed discussion in section IV.B), reveals again (Fig. 5.b) the parabolic profile of the diffusion front in a more sensitive way. We checked for intermediate crack speed that there is a continuous transition between quadratic profiles determined by height AFM images or phase images as well. It should be emphasized that no phase signal analog to that observed with soda-lime glasses as described in this paper was observed with alumino-silicate glasses where the alkali cation was lithium (see Fig. 2 c-d).

The value of parameter *B* (see equation 3) is thus obtained and variations of $v_c/B$ versus $v_c$ are then plotted for RH45% in figure 6 (full squares). Figure 6 reveals that $v_c/B$ is almost constant in a large domain of crack speeds. Similar experiments were done for different humidity rat: RH48% and RH58%. We note that $v_c/B$ slightly decreases for the highest values of $v_c$ ($v_c >$ 4nm/s). It is thus possible to get a first approximate value, $D_1^{45\%}$, of the expression $u_0^2 \cdot D_{eff}^{45\%}$: $D_1^{45\%} = 280 \pm 50 \text{nm}^2/\text{s}$ at 45%RH for $v_c <$ ~4nm/s. Same kind of treatment was done in the case of higher humidity rates 48% RH and 58% RH. Results are reported in Figure 6 (full triangles and diamonds). We got following results:

$$D_1^{48\%} \sim 1750 \pm 350 \text{nm}^2/\text{s} \tag{7}$$

and

$$D_1^{58\%} \sim 4100 \pm 1000 \text{nm}^2/\text{s} \tag{8}$$

for the lowest values of $v_c$.

### III.A.2 Influence of crack speed on position of the diffusion front along XX' axis

If the nucleation of nodules is related to a diffusion process then the position of diffusion front along the XX' axis with respect to the crack tip has to change with the mean crack speed. It can be predicted indeed that the distance from diffusion front to the crack tip (as measured on XX' axis) increases when crack speed decreases: That is effectively what is observed in case of RH = 45% (figure 7.a). The variation of $v_c$ was obtained by small variation of $K_I$ (due to slight changes in applied external force). $v_c$ was directly measured from AFM frames taken in same conditions (height mode, scan speed). In figure 7.b the variation of distance of diffusion front to the crack tip (as measured on XX' axis) is plotted versus the inverse of crack speed. The data are well fitted by a linear function. From equations (5) and (4), it is easy to see that the slope of curve $X = f(1/v_c)$ is equal to ($4 \cdot u_0^2 \cdot D_{eff}^{45\%}$). The calculated value of ($u_0^2 \cdot D_{eff}^{45\%}$) is then equal to $D_2^{45\%} = 160 \pm 80$ nm$^2$/s. This value is within the error bars of precedent measurement of $D_1^{45\%}$..

*III.A.3 Temporal and spatial variation of concentration: the case of 58%RH.*

Further studies were done in the case of lower crack speed (0.35nm/s) and 58%RH. In that case, thanks to the relatively high value of RH, diffusion process as detected by the nucleation of nodules is sufficiently fast even if the crack advance is strongly slowed down. We can thus study the local variation of concentration, $c$, of diffusing species versus time and position. $c$ is supposed to be proportional to the height, $H$, of nodules as measured in the AFM frames during that studied process of nucleation of nodules. We checked whether the studied process of the nucleation of nodules can be described by following equation characteristics of classical diffusion law [39]:

$$\frac{H(y)}{H_0} = 1 - erf\left[\frac{y}{2\cdot\sqrt{D_{eff}\cdot t}}\right] \tag{9}$$

That study was made on sequences of successive AFM height images: one example of these images is presented in figure 8. Two types of measurements were done:

α) At first we study the temporal variation of height of nodules (as that circled in white in figure 8). As seen in figure 9 the variation of the inverse of error function of ($1 - H/H_0$) versus the inverse of the square root of time is well fitted by a linear plot. From the slope we can calculate the value of $D_{eff}$: $D_{eff,\,t}^{58\%} = (16 \pm 4) \cdot 10^3$ nm$^2$/s.

β) Secondly we plot (figure 10) the inverse of error function of the argument ($1 - H/H_0$) as a function of $x$ for given values of $t$. For statistical reasons the calculation was done for all nodules included in the small zone of AFM frame (as in figure 8) delimited by the two white lines. The lower envelop[1] of data points (figure 10) corresponds to maximum height of nodules and is well fitted by a linear plot (full line in figure 10). From the slope of this linear fit the value of $D_{eff}$ can be calculated. The averaged value as obtained with a sequence of 10 images is equal to $D_{eff, X}^{58\%} = (19 \pm 5).10^3$ nm$^2$/s. At the accuracy of the measurements that value is equal to that given by the temporal determination. Consequently we took for $D_{eff}^{58\%}$ the mean value of these two measurements:

$$D_{eff}^{58\%} = (18 \pm 4).10^3 \text{ nm}^2/\text{s} \tag{10}$$

The consequence of these results is that an estimation of the parameter $u_0^2$ (supposed to be independent of the humidity rate) can be deduced by comparison between (8) and (10):

$$u_0^2 = 0.23 \tag{11}$$

It is now possible to calculate the value of $D_{eff}$ for the different values of RH as studied in this paper: the plot in figure 6 can now be graduated (right axis) with values of $D_{eff}$. The values of $D_{eff}$ for the regime of lower values of crack speed are summarized in table I.

### III.B. Higher speed regime of crack propagation: AFM phase images.

Typical phase and height images in the case of the "high" speed regime (crack speed higher than 3nm/s) are plotted in figure 11 (top images) in case of $v_c = 7$nm/s. The phase signal is the dephasing angle between the free oscillation of the AFM cantilever as measured far away from the sample surface and the signal measured during intermittent contact both at the nominal value of the oscillating frequency (arbitrary choice of origin of phase angle). As well known [40], the phase signal is a function of changes in the tip–sample energy dissipation. In figure 11 (graphs a2 and b2) the phase signal is lower in diffusion zone (inside the parabolic envelop line as mentioned above) than in the not-disturbed zone and corresponds, in the present experimental conditions, to an increase in dissipation the physical origin of which will be explained in section IV.B. The arrows indicate the position of crack line which is related to noticeable features: in figure 11.a1 (height mode) the crack corresponds to a deep and narrow region; phase signal (figure 11.a2) is tending to the typical value for a cantilever vibrating far away from the substrate as it is expected from a cantilever "flying" over the crack filled with standard surrounding gaseous atmosphere. On line (b) in phase image the crack is associated to a very narrow zone revealing a much lower phase signal. That increase of dissipation was associated [41] to the presence of a liquid condensate in the confined region between the two crack surfaces near the crack tip.

---

[1] The upper envelop corresponds to the genuine surface.

### III.C. AFM local investigation of the growth of nodules..

In order to get more accurate information about the origin and nature of the diffusion species we studied the growth of the nodules by AFM. For that purpose we used a much higher spatial resolution that allows us to work with relatively small times of acquisition (84s between two consecutive images). The investigated zone was chosen to be ahead of the crack tip at a distance of it along XX' equal to 1.8 μm and at a distance from crack line of 215nm. The conditions were RH=48%, $v_c$ = 0.21nm/s, $K_I$ = 0.35MPa.m$^{1/2}$. Sequence of AFM data was taken at a scan size of 1x1μm$^2$. Three numerically zoomed images from such a set of data are showed in figures 12.a to 12.c. In a first step the nucleation of nodules of very small size occurs all over the AFM scanned surface. Then the nodules grow in height and lateral radius before coalescing. An example of that process is shown in figure 13. The time evolution of portions of height profiles of ~200nm in length along YY' axis was investigated as shown in figure 13 (the studied profile is marked by the white line in figure 12.b). Two nodules (see arrows in figure 13) are first growing independently before coalescing to form one unique bigger nodule the height of which further increases. The time delay between the instant for the beginning of nucleation of first nodules and the coalescence is approximately equal to $\tau_1$ = 1050 ± 30s. Furthermore, thanks to a cross-correlation of AFM images, it was possible to correct the AFM data from residual AFM experimental drifts. We thus proved that the position of the nodules does not change with time.

We performed too a statistical study of such a phenomenon over the whole scanned surface of the sample (1μm$^2$). In figure 14 the mean radius of nodules as calculated for every AFM picture is plotted versus time in log-log scales. Two different regimes are clearly visible. The first one corresponds to a variation proportional to $t^{1/3}$ and the second one, for higher times, to a linear variation with $t$. The transition time between these two regimes is equal to $\tau_2 = 1000 \pm 50$s (same time origin as for $\tau_1$). This behaviour is well known and can be explained by the phenomenon of nucleation and coalescence of droplets (nodules) as in the case of the so-called "breath figures" [42, 43]. It was interpreted as the local condensation on nucleation sites of water molecules from the atmosphere to a liquid phase. Then, as the gaseous flux is constant, the volume of droplets/nodules increases in a first step linearly with time [44]. Thus the mean radius of them is a linear function of $t^{1/3}$. In the second step, as long as the one-nodule growth remains scale invariant, the growth of the entire pattern was shown [44] to be self-similar in time. Then, for the case of three-dimensional nodules condensing on a plane substrate, the exponent corresponding to a *mean* droplet, *averaged* over the pattern, was to be predicted to be equal to unity. This is indeed what is observed in figure 14 for time higher than 1000s. That latter transition time, $\tau_2$, is equal (at the accuracy of the experiments) to that, $\tau_1$, measured earlier in the case of the study of an individual nodule. It is a further proof of the validity of that interpretation of growing and coalescing nodules. It must be emphasized too that, for longer times, the self-similar regime should end when nodules grow so much that they can cover the whole surface by a wetting-like film [44]. This jamming limit is reached and a "thick" film is formed so the contrast in height images becomes very weak: this effect is very likely responsible for the absence of nodules in the close vicinity of the crack line (this effect is more visible when RH is higher: see figure 8).

To better understand the origin of these condensation patterns we measured by AFM the evolution of the contact angle versus time. A typical result is plotted in figure 15 in the case of the selected nodule (white line) in figure 12.b. The error in contact angle, $\theta$, was estimated to ~3° by calculating $tg(\theta)$ from mean roughness (0.1nm) and digitizing horizontal distance (2nm). The contact angle is increasing from ~ zero (complete wetting) to $18 \pm 3°$. This limit value was confirmed by a statistical study of the wetting angle at nodules over the whole area as that presented in figure 12 a-c.

This last study proves that nodules grow by a process similar to that involved in breath figures when liquid droplets are condensing on a cold surface. However it must be kept in mind that a similar phenomenon related to growth and coalescence of droplet-like particles can be observed in quite different circumstances as the growth of aggregates on a surface (silver on amorphous carbon). As water plays a major role in local chemistry at crack tip that condensation phenomenon evidenced by the observation of nucleation and growth of nodules is likely due to two conjugate effects: i) a local dewetting of the native thin layer continuous water (as evidenced by variation of contact angle) and ii) a condensation from the gaseous atmosphere. The aqueous liquid phase is likely enriched with ions and consequently has a higher viscosity.

### III.D. Ex-situ chemical micro-analysis

We need now to get information about the chemical nature of the diffusing species. Unfortunately very few information of this type at nanometer scale can actually be obtained from scanning probe experiments. In order to get deeper inside in the chemical nature of nodules we worked as presented now. Mean crack speed was chosen low enough (0.1 nm/s) and humidity high enough (48%) in order to allow an increase of the mean size of nodules with time. These experiments were done at room temperature and in the pure ($N_2$ +$H_2O$) atmosphere. Nodules of microscopic size laterally and in thickness were obtained. The sample was then withdrawn from the glove box, removed from the mechanical stage and stayed in standard atmosphere during a maximum of one hour before being put in high vacuum conditions. We verified that the nodules do not disappear during that stage. The (X,Y) surface of the sample was then studied in the vicinity of the crack tip. Hundred nanometers in thickness were removed by ionic sputtering in order to probe the inside of nodules. The variation of concentration of C, Na, Si, Mg and Ca over the analyzed surface (50µmx25µm) was studied. Inside the nodules (figure 16) it can be noted a much stronger, respectively lower, signal for sodium, resp. silicon, at nodules in contrast to the bulk glass. The concentration of other studied elements is constant all over the studied surface at the accuracy of the experiments. EPMA studies revealed similar features: a noticeable increase of concentration of sodium and a decrease in silicon signal at nodules. The study of the weak signals associated to Ca, Mg reveals no significant variations between the nodules and the bulk glass. Inside of nodules there is thus an enrichment in sodium when compared to the bulk glass.

## IV DISCUSSION

### IV.A. Summarizing the main features of the experimental results

The nucleation, growth and coalescence of nodules and the propagation of the front ("diffusion front") delimiting the "contaminated" (with nodules) and "uncontaminated" (without nodules) zones were observed at nanometer scale in the vicinity of the crack tip by AFM measurements in tapping mode in a carefully controlled surrounding atmosphere. Following points were outlined:

1. These experiments revealed that the "diffusion front" is connected to the nucleation processs of nodules. It has a parabolic profile with its symmetry's axis parallel to crack direction that is equivalent to a time axis because of the quasi-constant value of crack speed. These data can be well interpreted as due to a diffusion process.

2. We deduced from these data that the source of diffusing species is locating on (in the vicinity) of crack lips (in our 2D AFM study ; crack surfaces in case of general 3D conditions).

3. As stress field is drastically enhanced in the vicinity of crack tip, the role of stress gradient in the origin of the diffusion process is suspected.

4. Values of diffusion coefficient were measured at room temperature for different rates of relative humidity.

5. By further chemical microanalyses we revealed that nodules are characterized by a higher concentration of sodium element than in bulk glass. We then deduced that Na ions were migrating during the diffusion process.

6. AFM phase images showed that important changes in the tip–sample energy dissipation occurred. Water was proved to play a important role.

7. Complementary AFM studies revealed a nanometric process of nucleation and coalescence of the nodules connected to an increase of the contact angle.

We will now discuss following points : i) we will prove that diffusion of sodium ions will explain points 6) and 7) and high value for diffusion coefficients when compared with those of bulk soda-lime glass (point 4); ii) we will focus on the probable path of diffusing species (bulk or surface); iii) a tentative model of probable physical origin of the diffusion process will be proposed.

### IV.B. The role of the sodium cations in the observed diffusing process

SIMS, EPMA analyses proved that the ratio between concentrations in sodium and silicon is much higher in nodules than anywhere either on the studied surface far away from the crack line or in bulk glass. We did not note any detectable variation of concentration of other elements present in the bulk composition. These data are fully compatible with former results. It is indeed well known that even if soda-lime silicate glasses are very corrosion resistant at room temperature they can however react with water to form a superficial hydration layer. Numerous experiments established that the thickness of the corroded layer varies linearly with the square of leaching time in an initial stage and then linearly with time for longer periods of leaching (few hundred of hours at 90°C) [45]. A commonly accepted interpretation of this multi-step leaching behaviour is that the former is related to a diffusion-controlled process associated with the ion diffusion of $Na^+$ and/or $H^+$ (or $H_3O^+$), and the latter is caused by a chemical reaction at the surface of the glass. For the former process, a few possible mechanisms of diffusion have been proposed: Na self-diffusion from the inside to the outside of the glass [46,47] and interdiffusion of $Na^+$ in glass via protons or hydronium ions. However no real consensus has been reached on this matter. The mechanism most likely accepted concerned the interdiffusion of either hydrogen ions ($H^+$) or hydronium ($H_3O^+$) ions in solution with alkali ($Na^+$) ions in the glass. Such a process was first proposed by Doremus [46], then further experimentally studied by Lanford et al. [48]. These authors suggested that the most likely process is an ion exchange between $Na^+$ and $H_3O^+$. The same mechanism was proposed by Gehrke et al. [49-50] to explain the presence of a fatigue limit in crack growth in glasses: since $H_3O^+$ ions are bigger than $Na^+$ ions, a compressive zone is formed around the crack tip that causes the crack stop. This idea was reinforced by recent studies [7]. The important point in all these studies is that sodium ions in aqueous medium play a crucial role in corrosion studies of these glasses. As cations interdiffusion with hydronium ions is involved, control of acidic conditions is crucial especially in air atmosphere and/or stress corrosion experiments. The role of $CO_2$ for example in standard air is of huge importance in corrosion reactions with (alkali) silicate glasses surfaces. Experiments using AFM and X-ray photoelectron spectroscopy [51] revealed indeed that the combined presence of water vapor and carbon dioxide strongly reinforces the reactivity of the alkali silicate surface as compared with a bare water vapor atmosphere. In both cases the magnitude of the surface corrosion effect was found to increase in the order lithium << sodium << potassium. That last remark explains why the appearance of nodules due to alkali diffusion as observed in the present paper was only detected for silicate glasses where the alkali is sodium and not in case of lithium silicates (see figure 2). SIMS experiments proved that sodium content is higher inside nodules than in bulk glass.

The presence of local variation (in the nanometer range) of concentration of sodium ions near by the crack tip is the cause of local variations of surface tension and thus can explain the appearance of local aqueous droplets as it will be shown now. We indeed proved that the advance of diffusion front was related to a local and slight increase of wetting angle. On the opposite, far away from the crack tip, we did not observe any variation of the roughness of the glass surface: it is covered by a continuous thin layer of adsorbed water (the wetting angle, θ, is equal to zero) the thickness of which is in the range 10-60nm depending on surrounding humidity rate [52]. As it is well known the wetting angle is related to interfacial tensions by the Young–Laplace equation

$$\cos \theta = ( \gamma_{SG} - \gamma_{SL} ) / \gamma_{GL} \qquad (12)$$

where $\gamma_{XY}$ is the interfacial tension between phases X and Y. Subscript L is for liquid, G for gas and S for solid phase. At first it should be mentioned that because of the exchange in sodium and hydronium ions between bulk glass and water overlayer the concentration of $H_3O^+$ in the liquid phase decreases and then the solution becomes more basic. It is directly connected to the fact that sodium-containing glasses are regarded to be more basic than pure silica glasses [53]. It has been known for a long time that interfacial tension between liquid water and air, $\gamma_{GL}$, depends on the nature of the inorganic aqueous salt solution: $\gamma_{GL}$ increases when the pH of the liquid medium increases [54]. More recently molecular dynamics (MD) calculation [55] showed with more details that alkali cations are repelled from the liquid/gas interface in case of basic pH and, consequently, cause an *increase* of interfacial tension, $\gamma_{GL}$. As the alkali cations are repelled from the liquid/vapour interface their concentration near the solid/liquid interface should likely increase or remain constant. Thus $\gamma_{SL}$ should remain constant or slightly decrease. Experimental [56] and MD calculations [57] showed that there is small predominance of sodium ions at the interface of $Na_2O$-$3SiO_2$ glass and air, causing a little increase or no change in interface tension, $\gamma_{SG}$. Thus from the Young–Laplace equation it can be predicted that the contact angle should increase at the places where chemical heterogeneities (higher concentration of sodium cations) are present. That is indeed what is experimentally observed. Thus the observed nodules by AFM on height images and their growth are likely due to the local rupture of aqueous-based wetting film on the glass surface induced by chemical heterogeneities and the local condensation of water vapour. Far away and in front of the crack tip a stable wetting film is present: its equilibrium thickness is supposed to be in the range from 13 to 60nm [53] depending on the electrolyte concentration in the water film. The stability of such wetting film due to long-range van der Waals repulsive force [54] can be modified by external disturbances as the incoming of sodium cations causing local hydrophobic spots acting as nucleation centers for dewetting. As already shown [58] that circular droplets pattern – especially in the case of the thinnest (aqueous ) film – may develop by nucleation on more hydrophobic centers, ripen and merge. The nodule pattern is thus predicted to propagate with the diffusion front of sodium ions as it is observed in our experiments.

For the highest crack speeds the diffusion process is not rapid enough so that the height of nodules is too low to be detectable in the AFM height frames (Fig. 5.a and 11) acquired in tapping mode. However a negative and relatively strong signal is measurable in the phase frames in the "diffusion zone" (Fig. 5.b and Fig. 11). We indeed observe that in the diffusion zone (even if no nodules are present in height images) the phase angle (in algebric values) is much lower than the corresponding signal in diffusion zone which was itself lower than that in non-contaminated zone. It is known [41] that phase images in tapping-mode AFM are closely related to maps of dissipation: it was shown that if the amplitude of the cantilever is held constant, the sine of the phase angle of the driven vibration is then related to changes in the tip–sample energy dissipation. Calculation of dissipated energy are under progress [59] and will not detailed in this paper. The point of interest is here that a decrease in phase angle (in algebric values) correspond to an increase of dissipated energy during the tip-sample interaction. As mentioned in [60, 61] the principal source of energy dissipation during the tip-sample interaction is mainly the adhesion energy hysteresis. These adhesion forces can mainly be in the present experiments from two different origin [62]: capillary forces and acid/basic interactions [54]. In the first case the energy dissipation is related to the formation and rupture of a capillary neck between the tip and sample[63]. It is known that the capillary force is proportional to the liquid-vapor interfacial energy, $\gamma_{GL}$ [64] As earlier mentioned, $\gamma_{GL}$ is expected to increase due to an enrichment of the liquid-like phase in cations, we can thus easily explain the decrease in phase angle when the AFM tip is in intermittent contact in diffusion zone. On the other hand, the higher acid/basic interactions due to the already mentioned difference in acid-basic properties of both the silica AFM tip and sodium enriched zone on the substrate near the crack tip may also contribute to an increase of the dissipation in the diffusion zone. As mentioned by Fowkes et al. [54] work of adhesion between two silicate glasses (AFM tip and sample) is equal to the sum of two contributions : a van der Waals contribution and an acid-base contribution related to the ability to hydrogen-bonding. That last one is very likely increased due to more acid character of silica surface (AFM tip) and the more basic surface in Na-enriched glassy region.

### IV.C. Discussion on measured values for $Na^+$ diffusion coefficients. .

Values of inter-diffusion ($Na^+$ versus $H^+/H_3O^+$) coefficient for hydration experiments made in liquid conditions have been reported in the literature. If many studies were done by the leaching action of *liquid* water on glass substrates, at our knowledge, no study of diffusion processes was performed in gaseous water atmosphere. By using theoretical considerations and experimental observations, values for $Na^+$ diffusion coefficient and ratio between that coefficient and the $H_3O^+/H^+$ diffusion coefficient were found. For instance Lanford et al. [49] measured diffusion coefficient of $Na^+$ and $H_3O^+/H^+$ for the hydration of a soda-lime glass the composition of which is similar to that of the glass studied in the present paper. The sample was hydrated in distilled water at a temperature of 90°C. In that work glass plates were placed in distilled water containers for times ranging between few hours to hundreds of hours. Then the hydrogen content was profiled in depth by a resonant nuclear reaction method the spatial resolution of which is at least few micrometers. Typical values for diffusion coefficients are:

$D_{Na}^+ = 10$ nm$^2$/s and $D_{water} = 0.01$ nm$^2$/s at 90°C.

The high value of ratio between $D_{Na}^+$ and $D_{water}$ was confirmed by more recent experiments [34, 65, 66, 67]. Gehrke's work gave [51]: $D_{Na}^+ = 0,29$ nm$^2$/s for soda-lime glass rods of similar composition corroded in distilled water at 23°C. In these studies there was no feedback on pH but its value was supposed constant.

More recently leaching behaviour of sodium from small particles of soda–lime glass in acid solution was studied [68]. Two types of fine glass spheres were investigated: "small", respectively "large", ones with an averaged diameter of 50 (resp. 20) microns.

At 90°C:    $D_{Na}^{large} = 8.4$ nm$^2$/s   $D_{Na}^{small} = 27$ nm$^2$/s.

As the activation energy was measured in both cases we calculated sodium diffusion coefficients at a temperature of 25°C:

At 25°C:    $D_{Na}^{large} = 0.06$ nm$^2$/s   $D_{Na}^{small} = 0.6$ nm$^2$/s.

These values are very near of those obtained in former studies as in [51]. A slight increase of $D_{Na}^+$ with decreasing mean size of glass spheres is noticeable.

The first conclusion to be drawn from our measurements is that the values for $D_{Na}$ from literature are three orders of magnitude lower than those measured in the present study. It must be emphasized that i) these former studies were done by physical or chemical investigations at *macro*scopic or *micro*scopic levels and ii) no stress-gradient was applied during the measurement of diffusion coefficient. It must be noted too that the water content of silicate glasses has a strong influence of sodium diffusion coefficient. Schaeffer et al. [69] showed that increasing water content in $SiO_2$ considerably reduces the sodium mobility. More recently it was found that due to pre-annealing treatment in a wet furnace of a pure silica glass [70] or alkaline-earth boroaluminosilicate glass [71] an incorporation of water occurs in a near-surface region inducing a structural change which causes a much smaller sodium diffusivity in the near-surface region of the glass than in the bulk. By increasing the uptake of water in the structure of the studied silica glass it was however proved that that structural relaxation occurring in the near-surface region and leading to an initial rapid decrease in the sodium diffusivity is followed by a slow increase with the water content. Tomozawa et al. [72] showed that, for $Na_2O$-$3SiO_2$ glasses, with increasing water content sodium diffusion coefficient decreases initially to a minimum at 3-4wt % water and then increases. Thus this question of the influence of water content on mobility of sodium ions in silicate glasses is still controversial and no clear answer appears. However it should be concluded from literature that sodium diffusion inside silicate glasses with no or low content of sodium and without any applied mechanical stress is far below values measured in this paper.

The influence of the composition of silicate glass with sodium content was studied by Gehrke et al. [51]. An effective diffusion coefficient was measured by corrosion experiments of investigated glasses in distilled water at 23°C. Similar values for $D_{eff}$ were measured in soda-lime like glass (17 mol% in $Na_2O$) and for binary glasses ($Na_2O$ and $SiO_2$) with similar values of sodium content (12 mol% in $Na_2O$) as well: $D_{eff} \pm 0.3$ $nm^2/s$. However it was revealed an important increase of $D_{eff}$ with sodium content. Glasses made from 36% $Na_2O$ and 64% $SiO_2$ are characterized by $D_{eff}$ = 1.3 $10^3$ $nm^2/s$. That value is of the same order of magnitude as the one deduced from our experiments ($D_{eff}$ = 1 to 20 $10^3$ $nm^2/s$ for RH between 45% and 58%). The measured value of $D_{eff}$ by our local AFM measurements is thus fully compatible with the hypothesis of a zone, located in the surroundings of the crack tip, made of silicate glass with an enhanced concentration of sodium. An alternative explanation to the measured high value for $D_{eff}$ could be found in a local thermal effect. As diffusion processes are thermally activated processes it could be asked indeed whether the surroundings of crack tip do not support a non-negligible rise of temperature due to dissipation effects related to the advance of crack. It was indeed shown that *fast* cracks (near from Rayleigh speed) can generate heat at the crack tip causing a detectable increase of temperature [73,] and that a strong dissipation could bring the local crack tip temperature close to the melting temperature [74,75]. More recently experiments and simulations [76] of crack propagation in stainless steel at a speed in the range of mm/s (6 orders of magnitude higher than the crack speed in our experiments) revealed temperature rise of ~100°C. In case of a glass sample a maximum elevation of K was measured [77]. In our case the mean crack speed is in range of nm/s and temperature rise should be very small. Indeed we did not detect any rise in temperature during the advance of the crack in the present experiment. So an increase of diffusion constant due to a local increase of temperature as proposed by Brener et al. [78] for fast crack propagation is very likely excluded.

### IV.D. Probable path of diffusing species: bulk or surface?

We showed that the advance of the diffusion front during fracture –as observed on the free surface (stress plane) of the sample - is related to a migration of alkali ions. The source of diffusion was located at crack surfaces as discussed above. Two plausible scenarios for the migration path may be regarded. The first one corresponds to a *surface* migration of the diffusing species from the crack surfaces or from medium between them). The second scenario is related to a migration through the volume of material from crack surfaces to the outside of material: in this case transition between plane strain (inside the bulk material) and plane stress regions (near the free surface where AFM data are acquired) makes the observation of that phenomenon possible in these AFM experiments.

We will now show that the first scenario is very unlikely. It must be noted indeed that for usual values of relative humidity in our experiments (RH~45%) the glass surfaces are covered by an aqueous film the equilibrium thickness of which is in the range of few tenths of nanometers, depending on the electrolyte concentration in the water film [53]. Thus it could be possible that the migration of alkali species occurs in an aqueous liquid phase. In these conditions mobility of sodium cations in a bulk aqueous liquid phase is well known : $b_{Na} = 3.2 \ 10^{11} mN^{-1}/s$ [79]. Thanks to Stokes-Einstein relation we can calculate the diffusion coefficient of $Na^+$ ions in water, $D_{Na}^{water}$:

$$D_{Na}^{water} = b_{Na}.k_B.T$$

At 25°C we got $D_{Na}^{bulk\ water} = 1.33 \ 10^9 nm^2/s$, six orders of magnitude higher than the value we measured. As the water film on the surface of the sample has a limited thickness (in the range from 13 to 60 nm, depending on the electrolyte concentration in the water film) [53] effect of confinement on diffusion rate may be discussed. Recent Molecular Dynamics simulations [80] were done to investigate the influence of the size of cylindrical channel and its charge content on the diffusion of different ion. It was proved that for diameters as small as 1 nm no change or a slight decrease (maximum of one order of magnitude) in diffusion coefficient of alkali ions in liquid water are predicted. By using Stoke-Einstein relation with an sodium ionic radius, $R$, of $9.5.10^{-11}$m, it is possible to calculate the equivalent viscosity, $\eta_e$, of that hypothetic liquid-like surface layer the mobility of which would be observed in our experiments: we find $\eta_e = 1.9.10^{-3}$Pa.s. This value is much more lower than that we can deduced from our experimental value of diffusion coefficient: $2.10^3$Pa.s (~$2.10^4$Poises). This is a typical value for a soda-lime glass at a temperature around 1000°C. As a consequence the interpretation of our experiments by a surface migration of sodium in the surface aqueous film can be very likely ruled out. On the opposite, we note that the typical value of $D_{eff}$ we measured by our AFM experiments is very near from those obtained in case of diffusion of sodium through a bulk glass with an enriched content in sodium. We thus think that the migration of alkali ions we evidenced around the crack tip occurs through the bulk structure of glass. This effect is likely detected on the free surface of our sample -via a dewetting process- through the transition between strain-plane conditions in the bulk to stress-plane ones on the investigated surface.

### IV.E. The role of the stress gradient.

As the observed enhanced migration is localized in the vicinity of the propagating crack the enhanced value of stress around the crack tip is of major importance for the diffusion process.

It is known that mechanical stress has a great influence on diffusion processes. McAfee [81] studied the effect of stress upon the diffusion constant of gases (He, $H_2$ and other gases) in a borosilicate glass at room temperature. It was shown that an enhanced diffusion of helium (and hydrogen but with lower magnitude) is obtained in glass under high tensile stresses. On the opposite compression stresses had little or no effect on diffusion. This was interpreted by "the opening of channels" through the glass structure by tensile stress. Same kind of experiments were done later by Laska et al. [82] The same results were confirmed but the observed increase of the permeation rate of helium was interpreted as a decrease of "the effective thickness" of glass bulb.

The diffusion of water in silica glass was measured as a function of applied uniaxial stress at different temperatures by means of infrared absorption experiments [20, 83], nuclear reaction measurements [84]. These experimental data show that the water diffusion coefficient, $D_{H2O}$, into silica glasses is increased, respectively decreased, by an applied tensile, resp. compressive, stress in the case of temperature lower than 250°C. As ionic volume for $Na^+$ is ranged between those of $H^+$ and $H_3O^+$ ions the occurrence of a similar diffusion process for $Na^+$ ions in a soda-lime like glass can be predicted. A migration of sodium ions in the applied stress gradient was indeed deduced from electrical measurements [85] on bent soda-lime silicate glass plates studied at temperatures between 75°C and 120°C. More recently, Na peak emissions were detected [34] in vacuum by quadrupole mass spectrometry and surface ionization techniques from fractured soda lime glass and sodium trisilicate glass. These experiments revealed a fast diffusion (typical time in the range of milliseconds) of $Na^+$ in sodo-silicate glasses under high stress gradient. These observations are coherent with theoretical results from thermodynamics applied to chemo-elastic solid [86, 87] which allows the calculation of the diffusion coefficient versus the applied stress gradient. It was predicted that, for a given chemical composition and in case of interstitial species, the tensile stress reduces the diffusion potential. Thus the sodium ions are predicting to concentrate in the region of maximum tensile stress, i.e. in the nanometer surroundings ahead of the crack tip. This fast regime of fracture is observed when stress intensity factor is equal to the toughness, $K_I^c$, which is approximately equal to $0.8 MPa.m^{1/2}$. This value is "only" higher by a factor two when compared to the $K_I$ value in our experiments of very slow fracture: the main contribution to value of stress is indeed mostly due to the $r^{1/2}$ factor where $r$ is the distance from crack tip to the relevant point. Thus stress gradient have values of same order of magnitude in both regimes of slow and fast fracture. Therefore the fast migration of alkali ions from compressive zone to tensile one (at crack lips near crack tip) should likely occur in our AFM experiments during the first few tens of milliseconds. This effect is likely associated with an enhancement of the structural relaxation of glass near the fracture surfaces as already mentioned by Tomozawa et al. [73]. These authors showed indeed that surface structural relaxation is faster [23] than in the bulk and is accelerated by the presence of water vapor [23] and applied tensile stress [22]. Thus that millisecond range migration likely causes the presence of a reservoir of sodium ions in the nanometric vicinity of the crack lips which relaxes in a second step in a diffusion process with much higher time constants (few minutes) as evidenced in the AFM experiments detailed in this paper. As previously mentioned this slow relaxing diffusion process is associated with higher values of sodium and $H^+/H_3O^+$ interdiffusion coefficients than those measured with sodo silicate glasses with equivalent bulk composition but fully compatible with those measured in silicate glasses with an enhanced sodium content. This interdiffusion process has been experimentally evidenced thanks to the appearance of breath-figure like process, enhanced dissipation in AFM tip and substrate interaction and, in a latter step, the growth of nodules detectable in height mode.

Our observations of an increasing diffusion coefficient with humidity reinforced the conclusions of Agarwal et al. [23] of a faster relaxation in the presence of water vapour. It is also compatible with recent observations of Tian et al. [88] on silica glass. These authors showed indeed that the uptake of water in the structure of the studied silica glass causes a structural relaxation occurring in the near-surface region leading to an initial rapid decrease in the sodium diffusivity followed by a slow increase with the water content. Our results strengthen the theoretical concept of a near-surface structural relaxation [20]. We believe that the stress-gradient present at the vicinity of the crack tip is of major importance to enable the glass network to undergo a structural relaxation causing long-range structural changes, leading to modifications of bonding angles and distances and consequently enhancing mobility of alkali cations.

**IV.F. Hypothetic origin of the alkali dewetting centers : alkali diffusion channels?**

The experimental results of the present paper show that the sodium stress-enhanced diffusion is not homogeneously distributed over the surface near the crack tip. It is indeed localized on given "spots", the nucleation centers of the observed dewetting phenomenon. Such dewetting centers might be related to the sodium-rich channels as evidenced by molecular dynamics simulation [35] and inelastic neutron scattering experiments in sodium silicate melts and glasses [89]. These authors demonstrated that these channels and the related intermediate range order are frozen on cooling below the glass transition and serve as preferential ion conducting pathways in the relative immobile Si–O matrix. As it was proved that the formation of that preferential ion conducting pathways is not really affected by sodium oxide concentration that scenario deduced from results obtained with sodium-silicate glasses may be likely sustainable for soda-lime glasses. Thus the nucleation of the condensation pattern of the liquid-like phase might be occurring on the outgoing of that sodium ions channels at the free surface of the sample. A comment has to be made on the mean distance between these sodium channels: It was estimated from literature [90] to be of the order of 0.5-0.8nm which is significantly below the experimental value determined from our data (see for instance Fig. 13): 10nm. This discrepancy may be explained by a combined effect of, on one side, the limited lateral resolution in our AFM experiments (in the order of 10nm due to AFM tip radius of curvature) and, on the other side, the self-similarity of the dewetting process [44]. As the growth of the condensation pattern of the liquid-like phase is a self-similar process [44] the growth and coalescence steps are indeed only detectable when the mean distance between these nodules is higher than the AFM lateral resolution.

## V. CONCLUSION

AFM experiments were done in a gaseous atmosphere composed of a mixture of pure nitrogen/water vapors with soda lime silicate glasses. An enhanced diffusion of sodium ions in the stress-gradient field at the sub-micrometric vicinity of the crack tip was evidenced for relative humidity rate higher than 30%. The study of height images (in intermittent contact AFM mode) in the lowest crack speed regime and phase images in the fast crack regime clearly evidenced a slow diffusion process thanks to the observation of the growth of nodules in the vicinity of the crack tip and/or variation in phase signal. The source of the diffusion species was shown to be located at the nanometer vicinity of the crack surfaces. That phenomenon related to sodium and hydronium interdiffusion process was detected by AFM through the observation of several effects: a dewetting-like phenomenon evidenced by "breath figures"-like patterns, changes in the tip–sample energy dissipation as observed in phase imaging mode and growth of nodules in height images for lower crack speeds. Ex-situ experiments confirm that sodium ions are the diffusing species. These experimental results have been explained by a two-step process: i) a fast migration (typical time: few tens of milliseconds) of sodium ions towards the fracture surfaces as proposed by Langford et al. [34] , ii) a slow backwards diffusion of the cations as evidenced in these AFM experiments (typical time: few minutes). Diffusion coefficients of that relaxing process have been measured at room temperature for different relative humidity rate. The corresponding values were found to be much higher than those reported in literature with glasses of the same composition in similar conditions but were found to be compatible with those obtained for silicate glasses with a higher sodium content. The process of nucleation and growth of these nodules was locally investigated and explained by a phenomenon of local condensation of aqueous-based compound on nucleation centers formed by the outcoming of sodium cations from the bulk structure of the glass. The spatial heterogeneity of that nucleation process related to the sodium diffusion might be attributed to the presence of cationic channels as mentioned in literature [90].

Our results strengthen the theoretical concept of a near-surface structural relaxation [20]. We believe that the stress-gradient present at the vicinity of the crack tip is of major importance to enable the glass network to undergo a structural relaxation causing long-range structural changes, leading to modifications of bonding angles and distances and consequently enhancing mobility of alkali cations.

It must be emphasized too that both near-surface relaxation and the local outgoing of sodium from fracture surfaces and related aqueous chemistry make the indirect finding of the controversial damage nanocavities [5, 14] by *post-mortem* AFM investigation of the two corresponding fracture surfaces delicate.

Furthermore these studies reveal that surface crystallization of oxide glasses [90] or observation of nm-scale structures on air-exposed fractured surfaces of soda-lime silica [91] stems very likely from the ionic migration due to residual stress gradient and the related process of interdiffusion with $H^+/H_3O^+$ ions.

However much more work is needed in order to better understand the role of the chemical composition of glasses and their physical structure on these stress gradient enhanced migration of alkali cations. Another important point would be to understand the role of that sodium migration towards fracture surface on local chemical and physical properties as wetting, chemical reaction in the vicinity of the crack tip with standard air. Open questions too are the local measurement of exchange of electrical charges during the diffusion process and the reverse problem of knowing how an electrical field can control the local migration of cations. These questions will be addressed in the near future.

<div align="center">-=-=-=-</div>

**ACKNOWLEDGMENTS.**

This work was carried out at L.C.V.N., UMR CNRS 5587, University Montpellier 2, France. We thank M. George and L. Wondraczek for helpful discussions. We acknowledge E. Arnould. B. Delettre, J.-M. Fromental and R. Vialla for their valuable assistance. EPMA and SIMS experiments were done thanks to B. Boyer and C. Merlet. We are grateful to A. Mézy for a very efficient assistance.

# FIGURE CAPTIONS

Figure 1:

Sequence of three successive AFM height images in tapping mode for a soda-lime glass at 45% RH showing the advance of crack tip (mean crack speed is equal to 1.5nm/s). The vertical range is 10nm. The scan size is 5μm.

Figure 2:

Sequence of two successive AFM height (Figures 2.a and 2.b) and phase images (Figures 2.c and 2.d) in tapping mode for lithium alumino-silicate glass at 45% RH showing the advance of the crack tip. Data in (b) and (d) were taken eight minutes after (a) and (c). The vertical range is 5nm (resp. 40°) for height (resp. phase) images. The scan size is 1μm.

Figure 3:

AFM height image in tapping mode for a soda-lime glass at 45% RH. The white line corresponds to the best quadratic fit for delimiting zone between zones with and without nodules. Crack speed: 1.5 nm.s$^{-1}$. The vertical range is 10nm. The scan size is 5μm.

Figure 4:

Experimental (dots) contour of diffusion front for RH=45%, mean crack speed $v_c$ = 1.5nm/s. The full line is the best quadratic fit. Error bar is the mean standard deviation of five measurements.

Figure 5:

Height (a) and phase (b) images in case of soda-lime glass, RH=45% and $v_c$ = 7nm/s. Z range is 10nm (resp. 40°) for height (resp. phase) image.

Figure 6:

Variation of log ($v_c$/B) and effective diffusion coefficient, $D_{eff}$, versus mean crack speed, $v_c$. See text for definition of B. Lines are drawn as guides to the eyes. The experimental error (mean standard deviation of ten measurements) is of the same order of magnitude as the size of symbols.

Figure 7:

(a) Contours of diffusion fronts for different values of mean crack speed, $V_c$. The crack tip is at X=0. RH = 45%. Lines are drawn as guides to the eyes.

(b) Variation of distance from diffusion front to crack tip (full squares) versus the inverse of the crack speed. The full line represents the best linear fit (the correlation coefficient is equal to 0.997). Error bar is the mean standard deviation of five measurements.

Figure 8:

Height image in case of soda-lime glass, RH=58% $v_c$ = 0.35nm/s. The white circle delimits the zone where time evolution of nodule height will be studied (see fig. 9). The two white lines delimit the zone where spatial evolution of nodule height will be studied (see fig. 10). Z range = 50nm

Figure 9:

Time evolution of the inverse of error function the argument of which is a linear function of the nodule height, *H*, as measured in the zone delimited by the white circle in figure 8. $H_0$ is the nodule height at reference time. The full line represents the best linear fit (the correlation coefficient is equal to 0.994). Error bar is the mean standard deviation of five measurements.

Figure 10:

Spatial evolution (along the direction of propagation of the crack) of the inverse of error function the argument of which is a linear function of the nodule height, *H*, as measured in the zone delimited by the two white lines in figure 8. The full line represents the best linear fit of the lower envelope of data points (The correlation coefficient is equal to 0.922). Error bar is the mean standard deviation of five measurements.

Figure 11:

Height (upper left) and phase (upper right) images in case of soda-lime glass, RH=45% $v_c$ = 7nm/s. Height profiles (*a1* and *b1*) along lines at positions *a* and *b* respectively. Phase profiles (*a2* and *b2*) along lines at positions *a* and *b* respectively. These profiles are the average of ten successive scan lines. Scan size is 1μm and Z range for height (resp. phase) images is 5nm (resp. 30°). The arrows indicate the position of the crack line.

Figure 12:

Study of the growth of the nodules using AFM at a high spatial resolution. The time evolution of the profile of height along the white line is plotted in figure 13. RH=48%, $v_c$ = 0.21nm/s.

Figure 13:

Time evolution of height along profile marked by the white line in figure 12. The arrows are a guide for the eye to follow on the growth and coalescence of two nodules (coalescence at the bold line). RH=48%, $v_c$ = 0.21nm/s.

Figure 14:

Variation of mean radius of nodules –as averaged on AFM data as those in figure 12- versus time. RH=48%, $v_c$ = 0.21nm/s. Dotted lines are the best fitting curves: for higher values of radius the fitting function is linear (correlation coefficient equals to 0.989) and for lower values of radius cubic (correlation coefficient equals to 0.935). Error bar is the mean standard deviation of four measurements.

Figure 15:

Variation of contact angle of nodules versus time. RH=48%, $v_c$ = 0.21nm/s. The horizontal segment indicates the mean time of coalescence and its error bar (see figures 13, 14). The full line is drawn as a guide to the eyes. Error bar is the mean standard deviation of four measurements.

Figure 16:

SIMS data of the free surface of unbroken sample for different elements. The inside of nodules – thanks to ion sputtering- was studied.

**TABLE 1**

| Relative Humidity Rate (%) | $D_{eff}$ ($10^3$ nm$^2$/s) |
|---|---|
| 45 | 1.0 ± 0.2 |
| 48 | 7.6 ± 0.6 |
| 58 | 18 ± 1 |

Table I:

Variation of effective diffusion coefficient versus relative humidity rate.

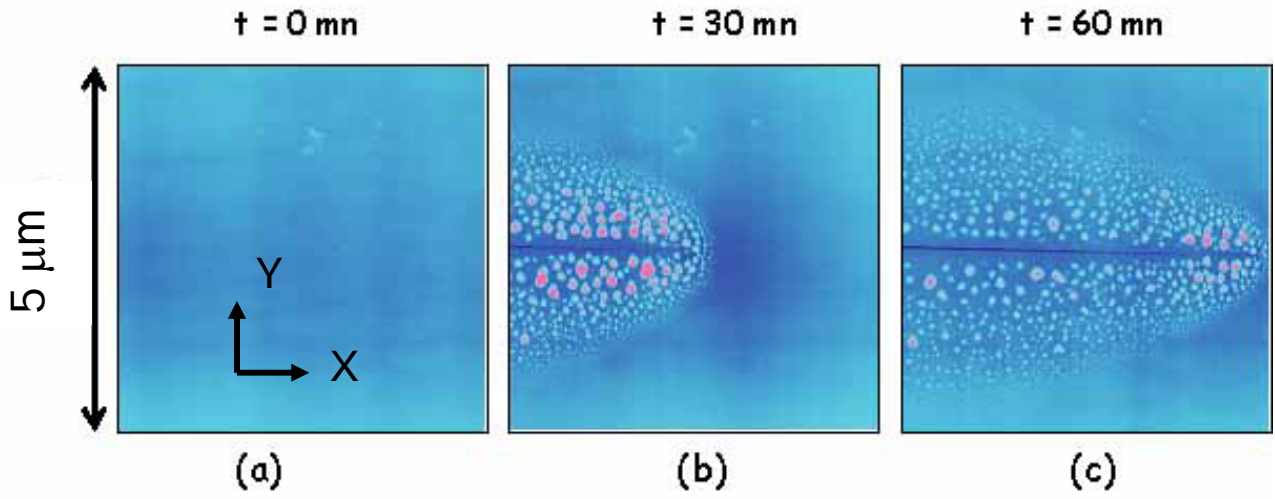

Figure 1

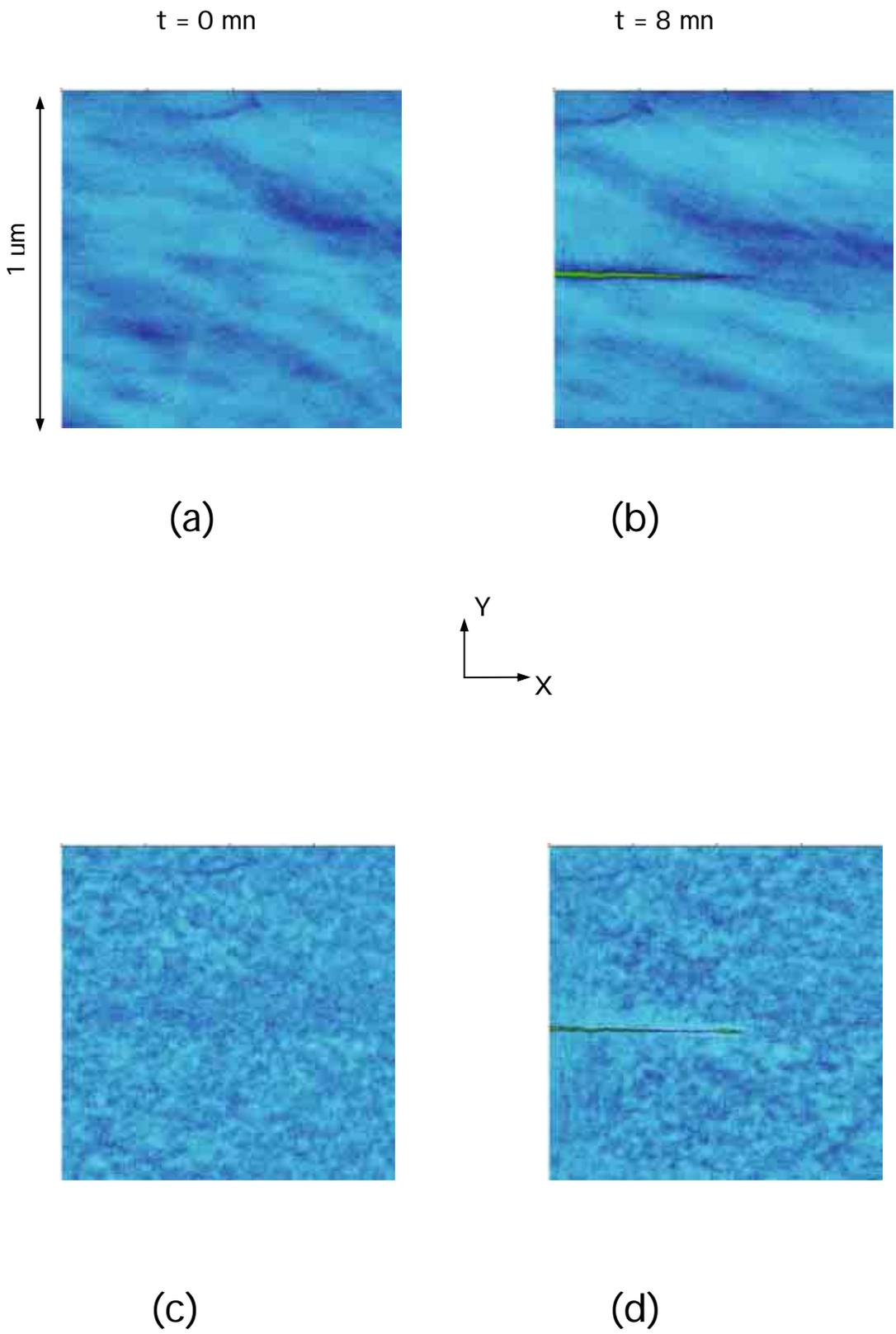

Figure. 2

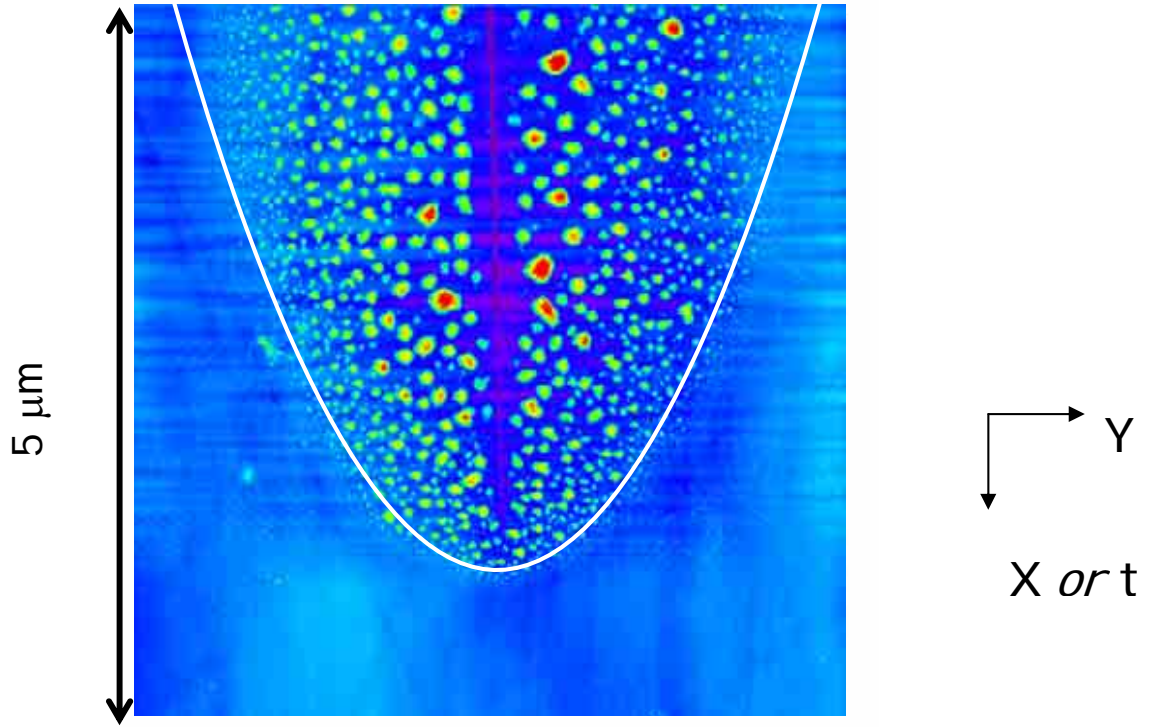

Figure 3

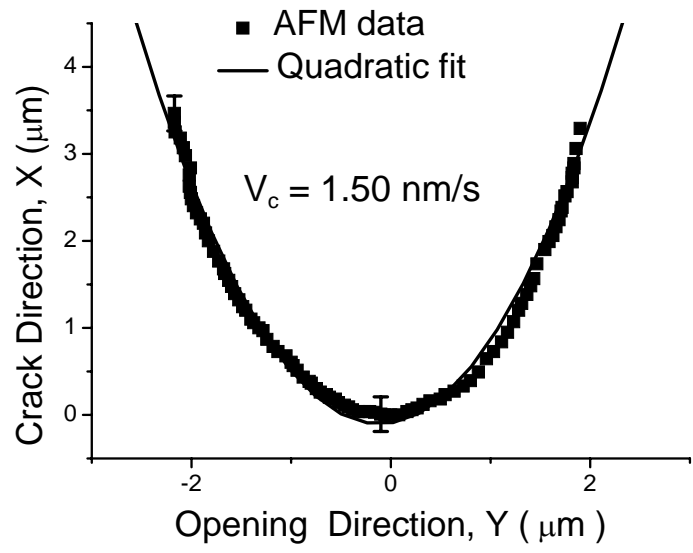

Figure 4

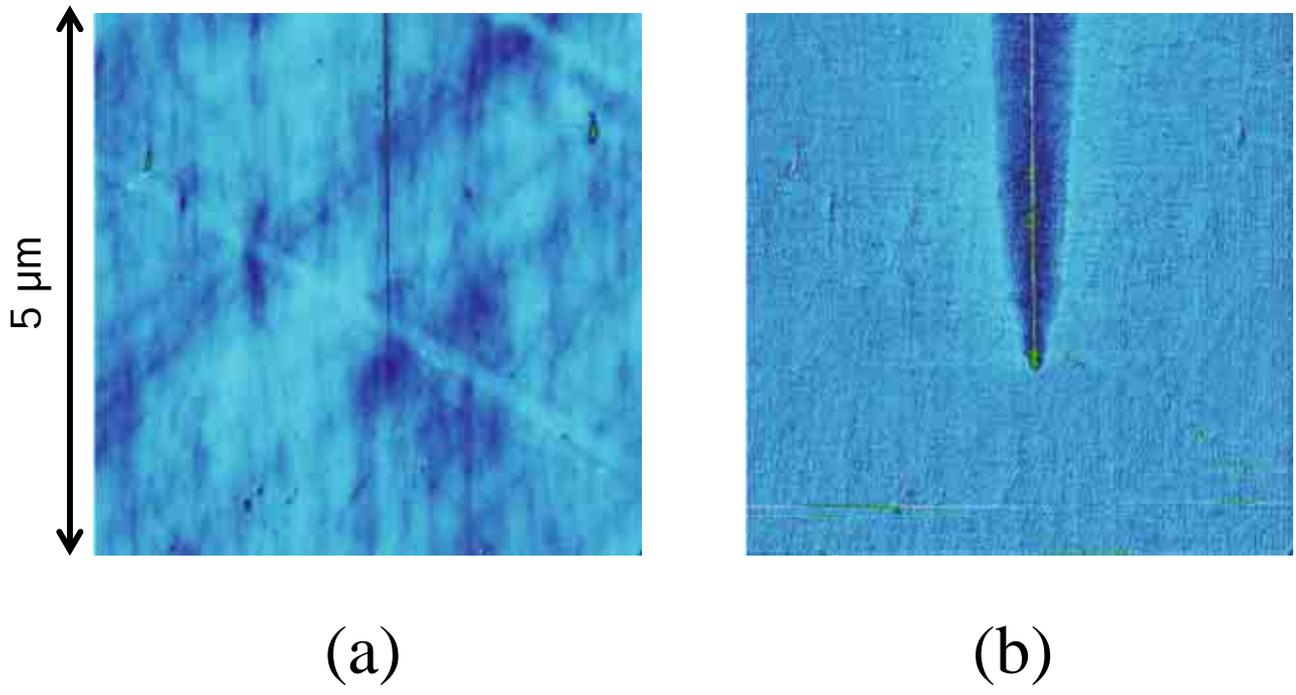

Figure 5

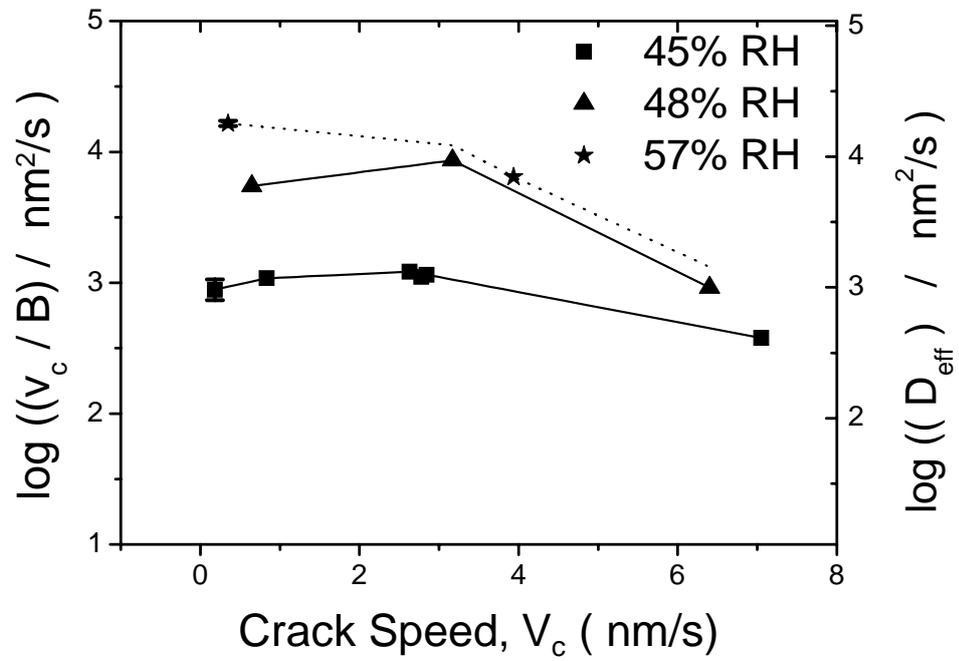

Figure 6

(a)

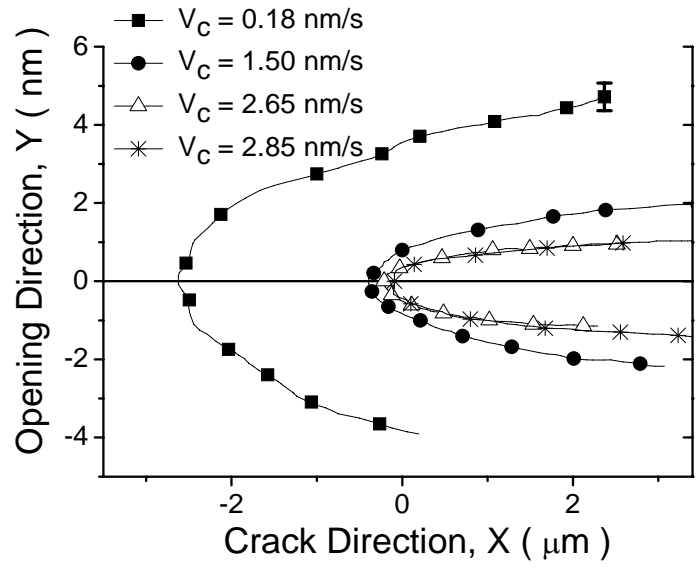

(b)

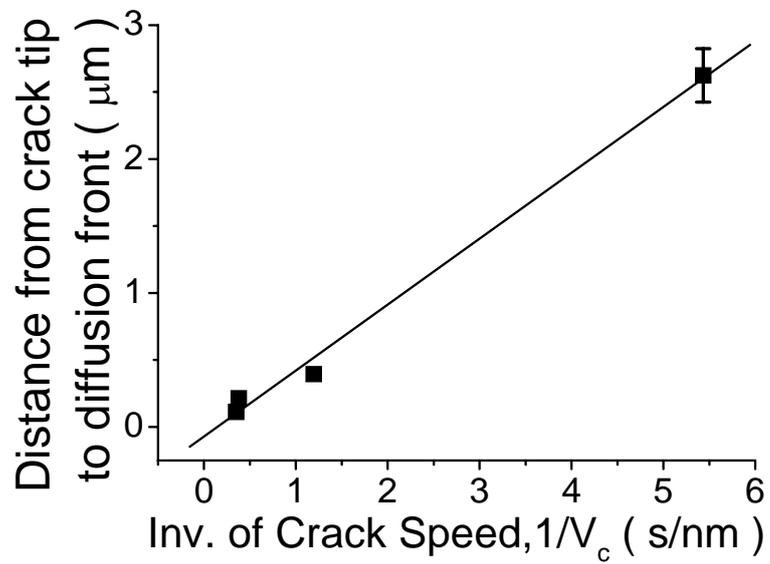

Figure 7

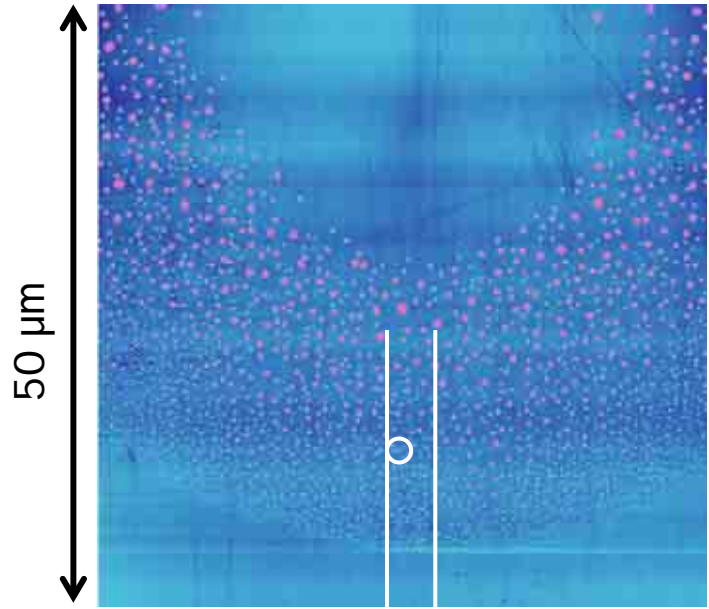

Figure 8

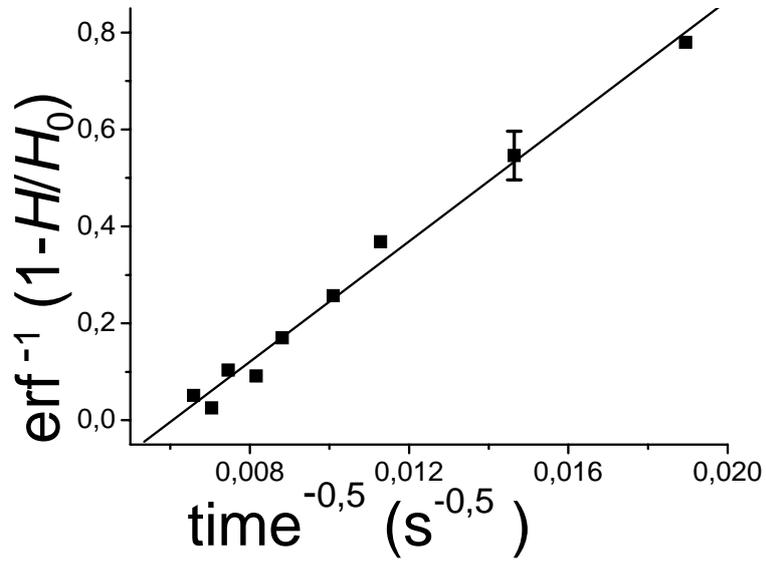

Figure 9

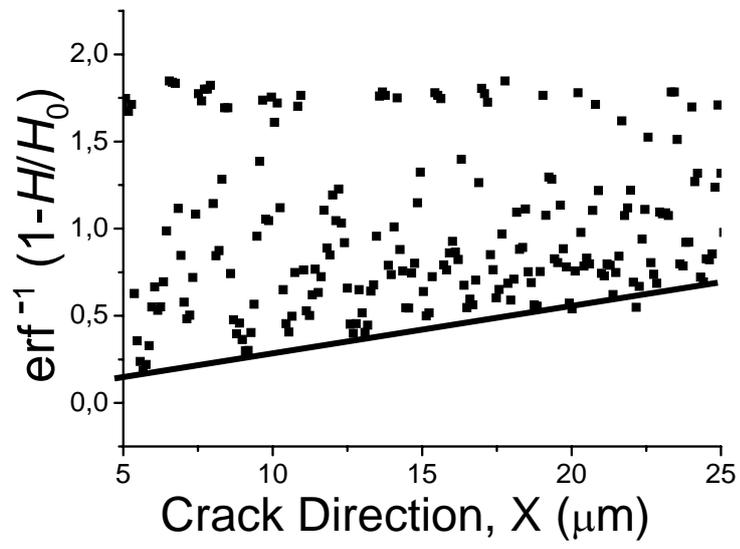

Figure 10

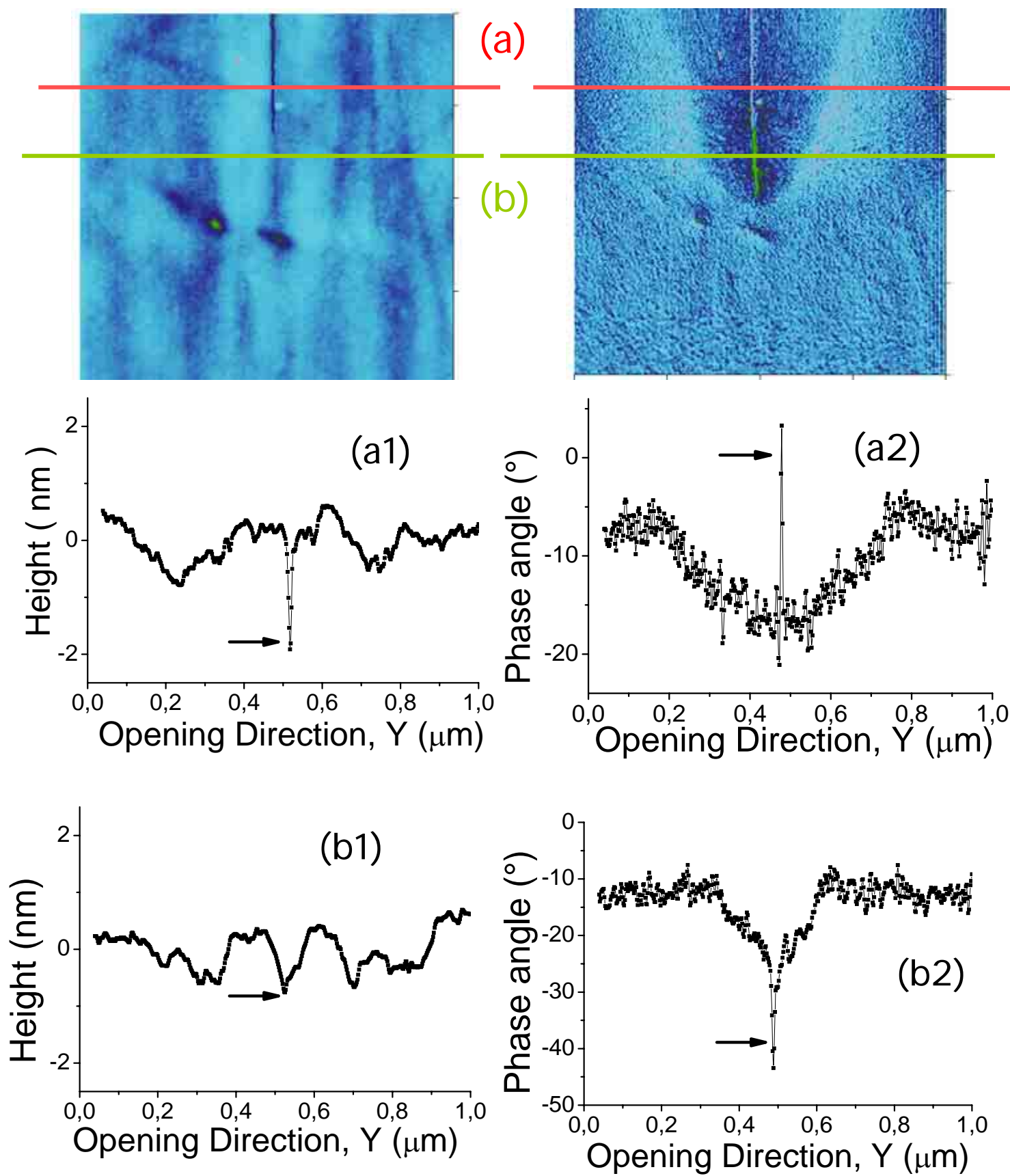

Figure 11

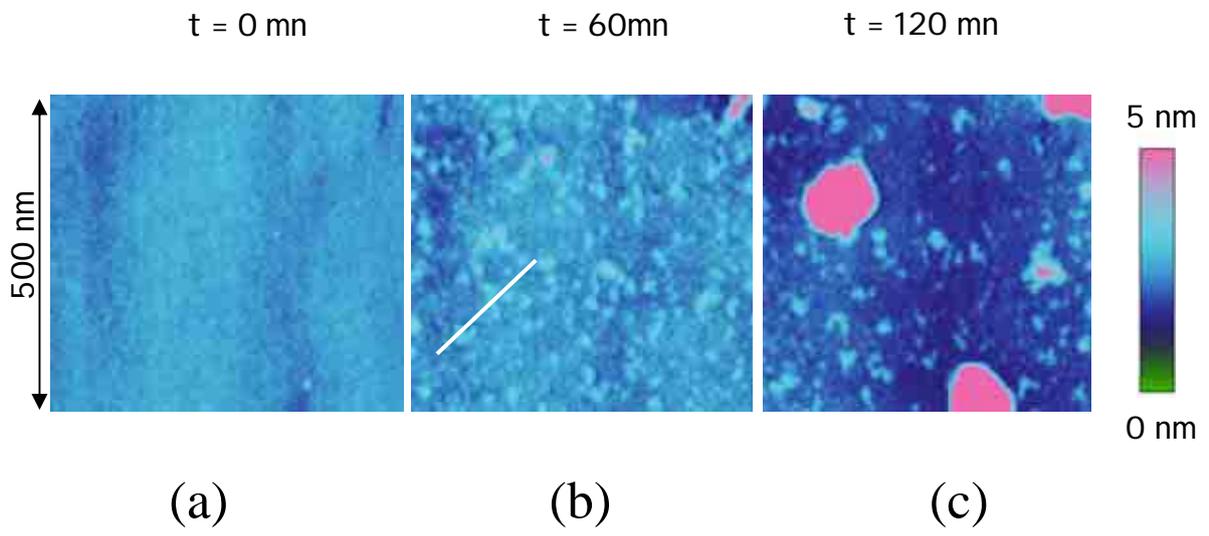

Figure 12

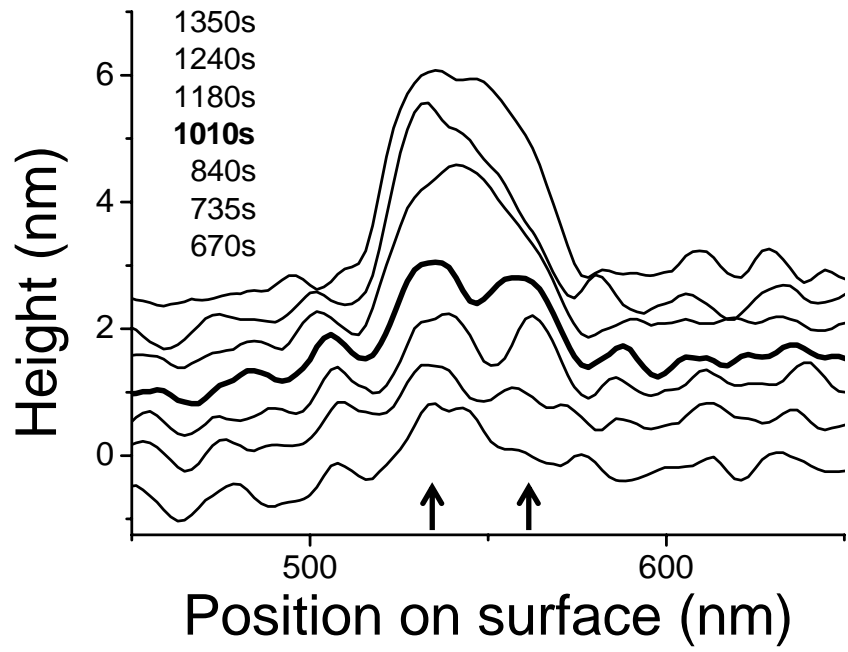

Figure 13

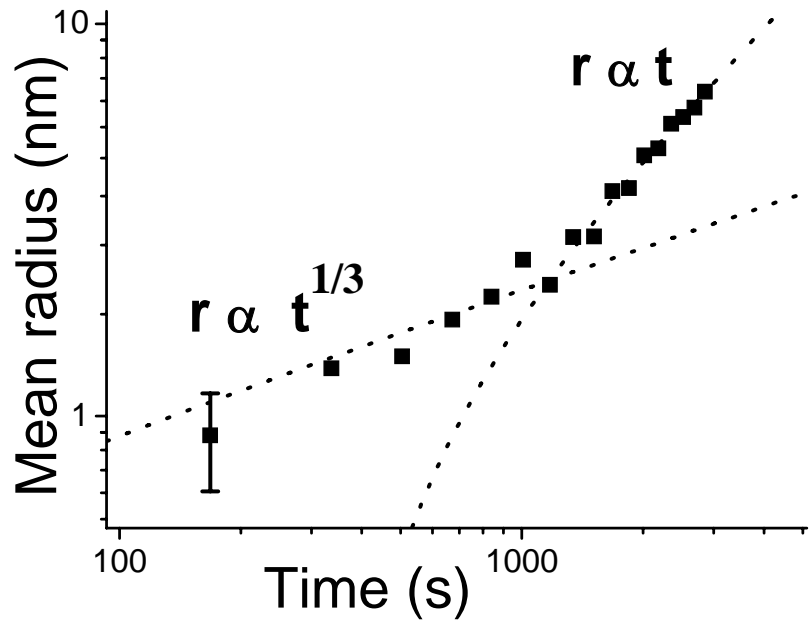

Figure 14

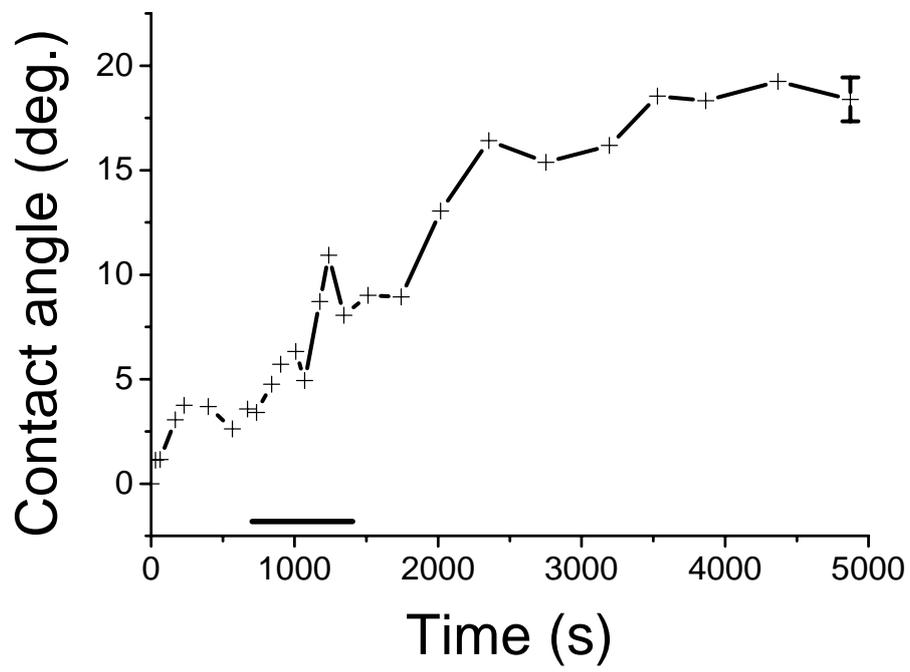

Figure 15

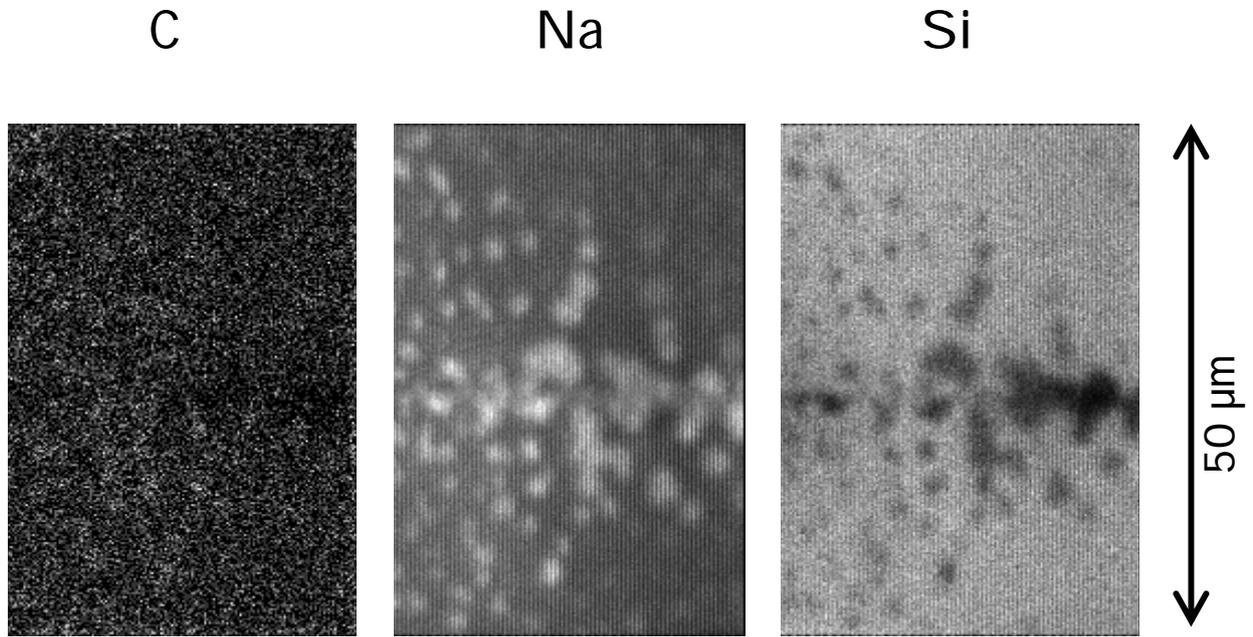

Figure 16